\documentclass[floatfix,aps,prd,eqsecnum,showpacs,amsmath,nofootinbib,preprintnumbers,twocolumn]{revtex4}

\usepackage{graphicx}

\def\laq{~\raise 0.4ex\hbox{$<$}\kern -0.8em\lower 0.62
ex\hbox{$\sim$}~}

\def\gaq{~\raise 0.4ex\hbox{$>$}\kern -0.7em\lower 0.62
ex\hbox{$\sim$}~}

\def\beq{\begin{equation}}
\def\eeq{\end{equation}}
\def\bea{\begin{eqnarray}}
\def\eea{\end{eqnarray}}
\def\bean{\begin{eqnarray*}}
\def\eean{\end{eqnarray*}}

\def \pa {\partial}
\def \ra {\rightarrow}

\def \Ms {M_{\rm s}}
\def \Mp {M_{\rm P}}

\def \vep {\varepsilon}

\begin{document}


\title{Macroscopic quantum tunneling and the ``cosmic" Josephson effect}

\author{A. Barone$^{1}$, M. Gasperini$^{2,3}$ and G. Rotoli$^{4}$}

\affiliation{$^{1}$Dipartimento di Scienze Fisiche, Universit\`a di Napoli Federico II, CNR-SPIN, Piazzale Tecchio 21, 80125 Napoli, Italy\\
$^{2}$Dipartimento di Fisica, Universit\`{a} di Bari, Via G. Amendola
173, 70126 Bari, Italy\\
$^{3}$Istituto Nazionale di Fisica Nucleare, Sezione di Bari, Bari, Italy\\
 $^{4}$Dipartimento di Ingegneria dell'Informazione,  Seconda Universit\`{a} di Napoli (SUN),  Via Roma 29, 81031 Aversa (CE), Italy}


\begin{abstract}
We discuss the possible influence of a cosmic magnetic  field on the macroscopic quantum tunneling process associated, in a cosmological context, to the decay of the ``false vacuum." We find a close analogy with the effects of an external magnetic field applied to a Josephson junction in the context of low-temperature/high-temperature superconducting devices. 
\end{abstract}

\vspace {1cm}~


\maketitle

\section {Introduction}
\label{sec1}
\setcounter{equation}{0}
It is well known that the interest in the Josephson effect \cite{1} lies in a variety of phenomena of paramount importance  for the deep physics involved, and for the many -- already realized or still  potential -- applications. In the field of superconductivity it represents a powerful probe of fundamental properties (such as, for instance, the unconventional symmetry of the order parameter in important classes of superconductors). However, the Josephson effect also offers an unparalleled tool to shed light on the intriguing underlying physics of many phenomena which go far beyond the realm of superconductivity. 

Many years ago the scientific community working in the field of superconductivity was greatly impressed by two papers  discussing the fate of the false vacuum \cite{cole1}, namely the stability/unstability of a physical system in a state which is not an absolute minimum of its energy density, and which is separated from the minimum by an effective potential barrier. It was shown, in those papers, that even if the state of the early Universe is too cold to activate a ``thermal" transition (via thermal fluctuations) to the lowest energy (i.e. ``true vacuum") state, a quantum decay from the false vacuum to the true vacuum may still be possible through a barrier penetration via macroscopic quantum tunneling (MQT). 

A few years later, the occurrence of the MQT in Josephson junctions \cite{5} became a subject of great interest.  For an excellent description of the experiments in this context, the reader is refereed to the paper by Clarke et al \cite{6}. The macroscopic quantum nature of the superconductive state, the role of the Josephson effect as a probe of MQT and of other quantum and macroscopic effects (such as energy level quantization and  macroscopic quantum coherence), produced a great development 
of both theoretical analyses and experimental confirmations. Unexpected results were also obtained, such as the first observation \cite{9} of MQT in high-temperature Josephson junctions (see Fig. 1), together with a clear observation of the energy-level-quantization effect \cite{10}. More recently, the influence of an external magnetic field applied to a Josephson junction was investigated, and it was found that the external field may play an important role in the activation of the MQT effect  \cite{11}.

Very recently, the decay of the false vacuum state in a cosmological context has attracted renewed interest, especially in view of its possible relevance in the process of tunneling among the many vacuum states of the string landscape \cite{13}. It seems therefore appropriate to point out that -- just like in the case of the MQT effect occurring in the context of a superconducting Josephson junction -- an external cosmic background field may have an important influence on the activation of the cosmological MQT effect, and on the effective decay rate of the false vacuum. The aim of this paper, in particular, is to estimate the effects of a cosmic magnetic field of a primordial origin, and to discuss its possible relevance for the false-vacuum $\to$ true-vacuum transition under the assumption that such  field satisfies the required energy density bounds imposed by inflation and backreaction (see e.g. \cite{15}).

The paper is organized as follows. In  Sec. \ref{Sec2} we briefly recall  the effect of an external magnetic field on the Josephson tunneling structures and, more generally, on the process of macroscopic quantum tunneling.  In Sec. \ref{Sec3} we generalize the original computation \cite{cole1} of the tunneling through the potential barrier for the decay of the false vacuum, including a cosmic background field of primordial origin. Our concluding remarks are finally presented in Sec. \ref{Sec4}. 


\section {External magnetic field and macroscopic quantum tunneling}
\label{Sec2}
\setcounter{equation}{0}

In the context of the Josephson effect, it is well known that the application of a magnetic-field $B$ to a Josephson junction produces a space modulation of the relative macroscopic phase \cite{1}. The effect of the applied magnetic-field on the related  MQT process was also a subject of deep researches, and stimulated a variety of interesting papers. Quite recently, in particular, 
the influence of such an external field on the so-called ``crossover" temperature $T_0$ has been explicitly computed \cite{11}, and it has been shown to be nonmonotonically dependent on the flux of the applied magnetic field $B$. 

\begin{figure}[t]
\centering
\includegraphics[width=87mm]{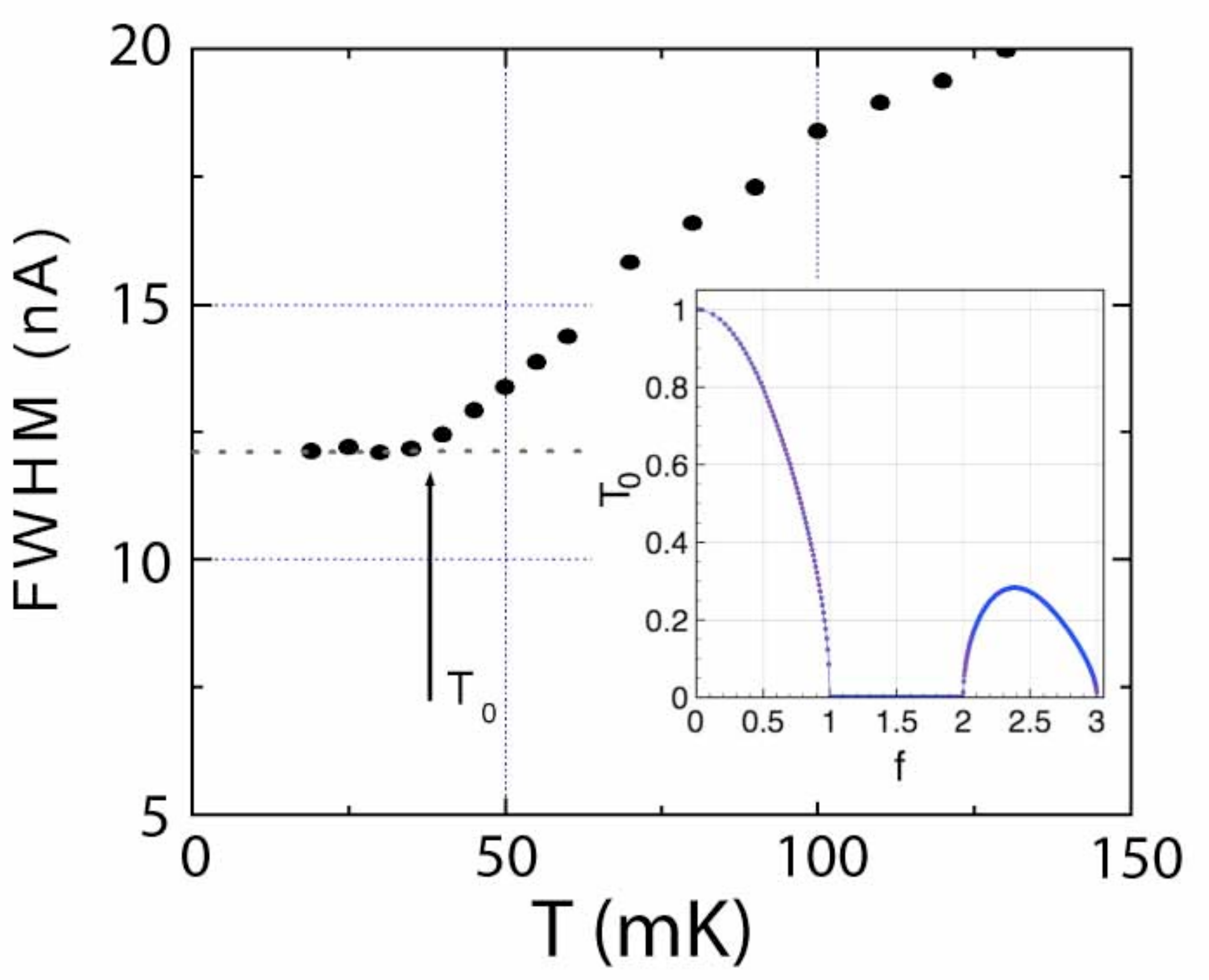} 
\caption{Details of full width half maximum of escape events distribution, as function of temperature, for a biepitaxial high temperature superconductivity Josephson junction in the experiment reported in \cite{9}. The  so-called crossover temperature $T_0$ is indicated by arrow. The plateau below $T_0$ indicate the setting of the condition of MQT (dotted line is just a guide for eyes). In the inset we show $T_0$ as a function of the flux $f(B)$ of the applied magnetic-field (see \cite{11} for the choice of units and the normalization of the plotted variables).}
\label{fig1}     
\end{figure}

Let us recall that  the crossover temperature $T_0$ is the (thermal bath) temperature at which quantum tunneling becomes the dominant 
activation mechanism of a superconducting Josephson junction: for $T<T_0$ the escape from the potential barrier is due to the MQT effect, while for $T>T_0$ it is mainly due to the effect of thermal fluctuations. An example of this is reported in Fig. \ref{fig1} where we plot the details of MQT transition in terms of the full width half maximum of escape event distributions for the YBa$_2$Cu$_3$O$_{7-x}$ (YBCO) biepitaxial Josephson junction of Ref. \cite{9}. The crossover temperature $T_0$, indicated by an arrow, separate the MQT regime from thermal regime.
In general the function $T_0=T_0(B)$ has  a complicated analytic form \cite{11} which depends on the junction parameters, and which tends to be periodic in the flux of the applied magnetic-field.  
Its general behavior is illustrated in the inset of Fig. \ref{fig1}.

It may be noted that, in a particular range of values of the applied field, 
there is no crossover and the junction remains classical until the zero temperature limit. This means that the external field may have a dramatic effect in controlling the ``shape" of the effective potential barrier and the overall efficiency of the MQT process -- at least in the context of  superconductive Josephson structures. What is important, for the purpose of this paper, is that the magnetic field produces a spatial variation of the phase whose main outcome is an increase in the tunneling rate \cite{11}, despite a decrease in the crossover temperature. 

Coming back  from the laboratory to the early universe, we may ask whether the above aspects of Josephson MQT can also be retrieved in a cosmological context (after all, the process of false vacuum decay discussed in \cite{cole1} is fully controlled by the tunneling probability). In particular, it would be an intriguing possibility if we could figure out an analogous behavior of the crossover temperature. However,  physical analogies deserve great interest but require, in general, also great caution. In our case it is probably inappropriate  to look for a cosmological effect exactly analogous to the one produced by the magnetic field on a Josephson junction, which influences so drastically the thermal vs quantum activation of the transition process. 

The reason is to be found in the substantial different nature of the  scalar field appearing in Coleman's model of vacuum decay and the (scalar) phase difference between two superconductors. Of course, it would be interesting to think of a ``cosmological" Josephson effect in which different regions of  the Universe have a phase difference in their (Wheeler-De Witt) wave functions, but this would represent a completely different problem from the one originally discussed in \cite{cole1}.

To find a possible cosmological analogous of the magnetic-field effects on MQT in Josephson structures we should consider, instead, the  scenario of false vacuum decay under the action of a cosmic background field nonminimally coupled to the scalar field of Coleman's model. 

We should recall, to this purpose, that the scalar dilaton field $\phi$ of superstring models is necessarily coupled to various antisymmetric tensor fields of second and higher ranks, including (in Type I and Heterotic superstrings) the electromagnetic field tensor $F_{\mu\nu}$ (see e.g. \cite{maurizio}). Such a coupling is typically of the form
\beq
\phi F^{\mu\nu}F_{\mu\nu}/(2M),
\label{21}
\eeq
where $M$ is an appropriate mass scale setting the effective coupling constant. Hence, if we are interested into MQT-type transitions among string vacua, we can identify Coleman's scalar with such a fundamental dilaton field, and revisit the original computation of the false vacuum decay taking into account the contribution of an external magnetic-field nonminimally coupled to $\phi$ according to Eq. (\ref{21}). This will be the subject of the next section.


\section{Magnetic-field perturbation in the semiclassical Coleman approach}
\label{Sec3}
\setcounter{equation}{0}

Let us consider the mechanism of false vacuum decay discussed in \cite{cole1}, based on a scalar field $\phi$ with a symmetric self-interaction  potential $U_+(\phi)=U_+(-\phi)$, and two minima at $\phi= \pm a$. Let us choose, in particular, the potential
\begin{equation}
U_{+}(\phi)=U_0\left( \phi ^{2}-a^{2}\right) ^{2}/a^4,
\label{31}
\end{equation}
where $U_0$ is the height  of the effective barrier separating the two 
minima of the unperturbed potential. 
Including the electromagnetic interaction (\ref{21}) the unperturbed scalar system is thus described by the following Lagrangian density: 
\beq
{\cal L}=  \pa_\mu \phi \pa^\mu \phi /2- U_{+}(\phi)
-{\phi}F^{\mu\nu}F_{\mu\nu}/(2M)
\label{32}
\eeq
(we are using units in which $\hbar=c=1$, and the action is dimensionless). Note that, for simplicity, we are working in the context of a flat space-time geometry. However, all subsequent computations could be easily generalised to the case of a curved cosmological geometry without changing the main results presented here. 

Following \cite{cole1}, let us now introduce a small breaking of the potential symmetry,
\beq
U_+ \ra U= U_+ +\varepsilon \left( \phi -a\right)/(2a),
\label{33}
\eeq
parametrised by the energy density parameter $\vep>0$ (which also represents, in the Coleman model without external electromagnetic interactions, the energy-density difference between the true and the false vacuum). The total effective potential thus becomes, in our case, 
\beq
V(\phi ) =U_{+}(\phi)+
\varepsilon \left( \phi -a\right)/(2a) +B^{2}\phi/M.
\label{34}
\eeq
Note that we have included only the contribute of the magnetic field $B$. Indeed, in a primordial cosmological context, one can safely neglect large-scale electric fields, assuming that they are rapidly dissipated by  the high conductivity of the primeval plasma \cite{16}. 

In order to obtain the exponent of the tunneling rate between the false and the true vacuum we need now the expression of the Euclidean action $S_E$ associated to the total effective potential (\ref{34}). Considering, as in \cite{cole1}, only corrections to $U_{+}(\phi)$ of the first order in $\vep$ and $B^2$ one obtains, in the ``thin-wall" approximation,
\begin{equation}
S_{E}=\pi ^{2}R^{3}S_{1}-2\pi^{2}\left[ \varepsilon 
 +\left(B^{2}a/{M}\right)\right] \left(R^{4}/{4}\right).
\label{35}
\end{equation}
Here $S_1=(8U_0)^{1/2}(a/3)$ is the ``wall" action, whose expression depends only on $U_{+}$, and $R$ is the radius of the four-dimensional hyperspherical ``bubble" of true vacuum  emerging, via MQT, from the initial false vacuum state. By minimizing the Eulidean action with respect to $R$ we obtain, for our model, 
$
R={3S_{1}}/(\varepsilon +aB^{2}/M).
$

It should be noted that above results are valid, within our approximations, provided the energy-density difference between the false and true vacuum is small with respect to the height of the unperturbed potential barrier, namely, for $\varepsilon +aB^{2}/M \ll 8U_{0}$. Assuming that such condition is satisfied, we can finally obtain the exponent of the tunneling rate -- also called ``bounce" coefficient, according to \cite{cole1} -- by evaluating the Euclidean action at the extremum value of $R$ given above. We find, as function of $U_{0}$:
\begin{equation}
S_{E}=\frac{27\pi ^{2}S_{1}^{4}}{2\left( \varepsilon +aB^{2}/{M}\right)^{3}}=
\frac{32}{3}\frac{\pi ^{2}a^{4}U_{0}^{2}}{\left( \varepsilon +aB^{2}/{M}\right)^{3}}.
\label{38}
\end{equation}
Using this equation, which represents the man result of this paper, 
we are now in the position of discussing the possible effects of a cosmic magnetic field on the decay of the false vacuum. 

Let us suppose, to this purpose, that the tunneling process occurs at a cosmological epoch  characterized by a value $H< \Mp$ of the Hubble parameter ($\Mp$ Planck mass), and localized around the end of the inflationary phase. Assuming that our scalar field is a dominant source of inflation we can estimate (using the Einstein equations at the given epoch) that $U_0\sim H^2M^2_P$; we 
can also give to the scalar field a typical value $\phi\sim a\sim \Mp$. A cosmic electromagnetic background field, on the other hand, can be naturally obtained by the inflationary amplification of the quantum vacuum fluctuations, and in this case we can evaluate its energy density by setting $B^2\sim H^4$ \cite{maurizio} (note that the energy bounds imposed by the process of quantum backreaction \cite{15} are satisfied, as $H^4 \ll \Mp^2 H^2$). 
Finally, a typical value of the photon-scalar coupling of Eq. (\ref{21}), for a scalar component of the multiplet of fundamental interactions, can be estimated to be of the order of $M\sim \Mp$.
In general, we can now consider two cases.

{\em A. Dominant magnetic energy.}
Let us use the above values of $U_0, \phi, a, B^2, M$, and let  us first assume that $\varepsilon \ll B^{2}a/M\sim H^4$, namely that the magnetic energy density dominates over the small  potential-energy asymmetry  $\vep$ used by  Coleman to trigger the MQT transition. In that case we have a fully magnetic activation of the MQT effect,  and the bounce  coefficient (\ref{38}) becomes 
\begin{equation}
S^{\rm mag}_E=\frac{32}{3}\frac{\pi ^{2}a^{4}U_{0}^{2}}{\left(aB^{2}/{M}\right)^{3}} \sim \frac{32\pi^2 }{3}\left(\frac{\Mp}{H}\right)^8.
\label{39}
\end{equation}
The approximation condition (\ref{38}) becomes $ H^2 \ll \Mp^2$, and is always satisfied as the inflation scale $H$ is always smaller than the Planck scale $\Mp$. This implies, according to Eq. (\ref{39}), 
$S^{\rm mag}_E \gg 1$, namely a very small value of the Bounce coefficient.

We may recall, in particular, that in the standard inflationary scenario the typical inflation scale must satisfy the constraint $H \laq 10^{-5}\Mp$ (following from the inflationary production of a primordial background of gravitational radiation, see e.g. \cite{17}). One thus obtains, in this context, $S^{\rm mag}_E \gaq 10^{40}$. Such a lower bound can be somewhat relaxed in a string cosmology context, where the inflationary scale is allowed to be as high as the string mass scale $\Ms$, i.e. $H \laq \Ms \sim 10^{-1} \Mp$ (see e.g. \cite{maurizio}). 
In that case one obtains $S^{\rm mag}_E \gaq 10^{8}$.

Larger values of the bounce coefficient can be obtained, however, if the Coleman term proportional to $\vep$ dominates the potential-energy  asymmetry. In that case the magnetic field $B$ may enhance the probability of tunneling transition, playing a role very similar to that played in the case of a microscopic Josephson junction.

{\em B. Subdominant magnetic energy.}
Assuming that $\varepsilon \gg B^{2}a/M\sim H^4$ the bouncing coefficient (\ref{38}) becomes
\begin{equation}
S_E\simeq \frac{32}{3}\frac{\pi ^{2}a^{4}U_{0}^{2}}{\varepsilon ^{3}}\left( 1-
\frac{3B^{2}a}{\varepsilon M}\right) =S_{E}^0\left( 1-\frac{3B^{2}a}{
\varepsilon M}\right) ,
\label{310}
\end{equation}
where $S_E^0$ denotes the Coleman result without magnetic-field contributions \cite{cole1}. Using the previous values for the parameters of our cosmological scenario we have $S_E^0 \sim 32\pi^2 \Mp^8 H^4/3\vep^3$,
and we obtain that the variation of the bounce coefficient induced by the magnetic-field is given by 
\beq
\frac {\Delta S_E}{S_E^0} \equiv {S_E-S_E^0\over S_E^0} \simeq - 3 
\frac {H^4}{\epsilon } \ll 1.
\label{312}
\eeq
The corresponding enhancement of the escape rate is, in first approximation,
\begin{equation}
\frac{\Delta\Gamma}{\Gamma _{0}}=e^{-\left( S_E-S_{E}^0\right)}-1\simeq 3 S_E^0 \frac {H^4}{\epsilon } 
\sim\left(\frac{M^2_P H^2}{\varepsilon}\right)^4.
\label{313}
\end{equation}

It is interesting to note that the condition for validity of our approximation, namely a small energy perturbation with respect to the height of the unperturbed barrier, implies in this case $\vep \ll 8 U_0 \sim  8\Mp^2 H^2$ (and is still compatible with the previous assumption $ \vep \gg H^4$). It follows that 
$\Delta\Gamma/\Gamma_0 \gg 1$. Hence, for $H^4 \ll \vep \ll \Mp^2 H^2$, even a small variation of the bounce, induced by the cosmic magnetic background according to Eq. (\ref{312}), may result into a large variation of the escape rate for the decay of the false vacuum state. 

\section{Conclusion}
\label{Sec4}
\setcounter{equation}{0}

The decay of the false vacuum discussed in two pioneer papers by Coleman and, Callan and Coleman \cite{cole1}
is based on a process of MQT very similar to the one occurring in the context of superconductive Josephson structures \cite{1}. The behavior of such superconducting devices is strongly influenced by the presence of an external magnetic field. In this paper we have shown that also in a cosmological context the false vacuum decay may be affected by the contribution of a cosmic magnetic field. 

Working in the thin-wall approximation, to first order in the magnetic energy density, we have found that a magnetic enhancement of the MQT escape rate is possible, and consistent with a realistic cosmological scenario satisfying the conventional  bounds imposed by inflation and quantum backreation.

The results obtained in this paper can be directly applied, in particular, to the process of tunneling among the vacuum states of the string landscape \cite{13}, and can be easily generalised to take into account the contribution of cosmic background fields of nonmagnetic origin. A more detailed investigation, taking into account other possible contributions to the tunneling rate in the Planckian regime, is postponed to a future paper.

\section*{ACKNOWLEDGMENTS}
The authors warmly thank A. J. Leggett for fruitful suggestions.
The work has been partially supported by the EC STREP Project MIDAS.

\end{document}